\documentclass[aps,prl,reprint,groupedaddress]{revtex4-2}

\usepackage{textcomp}
\usepackage{graphicx} 
\usepackage[tbtags]{amsmath}
\usepackage{physics}
\usepackage{multirow}
\usepackage{color, soul} 
\usepackage{enumitem}

\usepackage{hyperref}% add hypertext capabilities

\begin{document}

%\preprint{}

\title{Terbium under High Pressure: First-Principles Dynamical Mean-Field Theory Study}


\author{Wenjun Ding}
\email{wding@uab.edu}
\affiliation{Department of Physics, University of Alabama at Birmingham, Birmingham, Alabama 35294, USA}
\author{Yogesh K. Vohra}
\affiliation{Department of Physics, University of Alabama at Birmingham, Birmingham, Alabama 35294, USA}
\author{Cheng-Chien Chen}
\email{chencc@uab.edu}
\affiliation{Department of Physics, University of Alabama at Birmingham, Birmingham, Alabama 35294, USA}

\date{\today}

\begin{abstract}
Elemental rare-earth metals provide a playground for studying novel electron correlation effects and complex magnetism. However, {\it{ab initio}} simulations of these systems remain challenging. Here, we employ fully charge self-consistent density functional theory and dynamical mean-field theory (DFT+DMFT) to investigate terbium (Tb) metal under pressure. We show that Tb exhibits a strong band renormalization due to correlation effects, with the calculated electron density of states in good agreement with the experiments. At higher pressures, the correlated electronic structures persist but with modulation in the Hubbard gap, highlighting the tunability of effective Coulomb interactions and kinetic energies. Our DFT+DMFT calculations further indicate a ferromagnetic ground state of Tb at low pressure and low temperature, as well as a transition from ferromagnetism to paramagnetism at elevated temperatures. These {\it{ab initio}} results also align with the experiments. Our study paves the way for exploring heavy lanthanides via advanced first-principles simulations.
\end{abstract}

%\keywords{density functional theory, dynamical mean-field theory, terbium, rare-earth metal, high pressure}

\maketitle

{\it Introduction --}
Rare-earth materials have been rich platforms for exploring novel correlated phenomena, such as valence fluctuations, heavy fermions, Kondo physics, and giant magnetostrictions~\cite{Lawrence81RPP, Ernst11N, Aynajian12N, Chen18PRL, Jiles94JPD}. As a key controllable parameter, externally applied pressure offers the ability to tune the underlying physical properties and/or induce phase transitions. In the prototypical volume collapse example, the light lanthanide metal cerium (Ce) undergoes an isostructural transition at a pressure $P \sim 0.8$ GPa, where the volume is significantly reduced by $\sim 15\%$. The transition is accompanied by drastic band structure variations, which have been widely interpreted as the competition between Mott physics of interacting $4f$ electrons and their Kondo coupling to itinerant $5d/6s$ bands~\cite{allen1982kondo, lavagna1982volume, Haule05PRL, rueff2006probing, lipp2012x, Amadon14PRB, Chen19JPCL}. 

In heavy lanthanide metals like terbium (Tb), increasing pressure typically induces a common series of structural transitions, from the hexagonal close-packed (hcp) to a samarium-like ($\alpha$-Sm) phase, and then to a double hexagonal close-packed (dhcp) phase~\cite{samudrala2013structural,mcmahon2019structure}.
These phases exhibit a complex magnetic phase diagram with pressure and temperature.
Specifically, Tb in the low-pressure hcp phase is paramagnetic (PM) at room temperature but transitions to a helical antiferromagnetic (helical-AFM) phase at 229 K, and then to a ferromagnetic (FM) state below 221 K~\cite{Koehler63JAP, Koehler65JAP, Dietrich67PR}. 
Ferromagnetism is suppressed at higher pressure, and eventually the $\alpha$-Sm and dhcp phases exhibit low-temperature AFM orders, although their precise spin configurations are elusive~\cite{Matthew23JMMM, Kozlenko21PRM, Mito21PRB}. Tb is a relatively simple system among the heavy lanthanides, and it offers the opportunity to study the interplay of charge, spin, and lattice variables, where theory and computation can provide further insights to their correlated behavior and emergent magnetism.

The traditional workforce of density functional theory (DFT)~\cite{Hohenberg64PR, Kohn65PR} with simplified energy functionals highly underestimates correlation effects, so beyond-DFT methods for interacting $4f$ electrons are required. Extensive first-principles simulations employing the open-core approximations~\cite{Kozlenko21PRM}, Hubbard-I approximations~\cite{McMahon19PRB, Locht16PRB, Lebegue06JPCM}, or the Hubbard $U$ corrections (DFT+U)~\cite{Logan24JPCM} have been made to study Tb under pressure, achieving overall good theory-experiment agreements in different aspects~\cite{Kozlenko21PRM, McMahon19PRB, Locht16PRB, Lebegue06JPCM, Lang81JPFMP, Logan24JPCM}. However, these methods are semi-empirical or neglecting the coupling between correlated orbitals and effective bath sites, which may limit their predictive capabilities in highly compressed correlated electrons. Fully {\it ab initio} quantum many-body calculations beyond static mean-field theory remain challenging.

In this work, we perform fully charge self-consistent density functional theory with dynamical mean-field theory (DFT+DMFT) calculations to study Tb metal. Our {\it ab initio} electronic structures exhibit sharp coherent quasiparticle peaks associated with $4f$ orbitals, where the occupied and unoccupied electron density of states (DOS) near the Fermi level resemble the correlated lower and upper Hubbard bands, respectively. Our ambient-pressure spectra agree quantitatively well with spectroscopic measurements, and the main theoretical features are robust against variations in the computation schemes and interaction parameters. At high pressures, the correlated spectra become increasingly broadened with a reduced Hubbard gap. These predictions are consistent with the expectation that the effective Coulomb interactions are increasingly screened by pressure-enhanced charge hybridization. This modulation of the electronic structure also demonstrates a high-pressure tunability of the relative strength between electron interaction and kinetic energy. Finally, our calculations indicate a low-temperature FM state in the hcp phase, and a suppression of the FM moments by increasing temperature. 
These results agree with the experiments, showing the potential of predictive modeling using fully charge self-consistent DFT+DMFT calculations to explore Tb and other heavy lanthanides under pressure.

{\it Methods --}
An accurate treatment of correlation effects is critical for describing rare-earth materials. To this end, we employ the dynamical mean-field theory (DMFT)~\cite{Kotliar96RMP, Kotliar06RMP}, which is a state-of-the-art numerical technique for strongly correlated electrons. In DMFT, the original interacting lattice problem is mapped to a quantum impurity embedded in a bath environment. An impurity solver is then utilized to obtain the impurity self-energy~\cite{Werner06PRL, Haule07PRB, Gull11RMP}, which in turn is approximated as the self-energy of the original lattice model. The procedure needs to be repeated until the self-energy converges, representing the first self-consistent loop in our calculations. 

The (non-interacting) tight-binding parameters of the lattice Hamiltonian provide material-specific information on the underlying atoms, orbitals, and crystal structures. The tight-binding parameters are obtained from DFT, while the correlation strength is controlled by the on-site Hubbard interaction $U$ treated by DMFT. Together, these lead to a DFT+DMFT method~\cite{Lichtenstein01PRL, Lechermann06PRB, Amadon08PRB}. Instead of the one-shot DMFT commonly employed in the literature, here we use a fully charge self-consistent DFT+DMFT approach, where the resulting DMFT charge density is fed back to DFT calculations. The procedure is repeated until the charge density between DFT and DMFT converges, representing the second self-consistent loop. Figure A1 in the Appendix shows a schematic workflow of the calculations.

The DFT calculations are performed using the WIEN2k package~\cite{Blaha20JCP}, an all-electron full-potential linearized augmented-plane-wave (LAPW) method for describing both valence and core electrons. The DMFT calculations are based on the EDMFTF software~\cite{Haule10PRB, Blaha20JCP}, which provides quantum impurity solvers and integrates with WIEN2k to achieve full charge self-consistency. The input parameters are discussed in the Supplemental Material~\cite{SM}, which also contains additional calculation results discussed later. It is noted that not all calculations can be stabilized in the current DFT+DMFT implementation, partly due to the larger configurational space of 4$f$ orbitals in heavy lanthanides. Figure A2 in the Appendix summarizes potential convergence issues and our solutions, based on systematic and comprehensive tests.

\begin{figure}
\includegraphics[width=0.85\columnwidth]{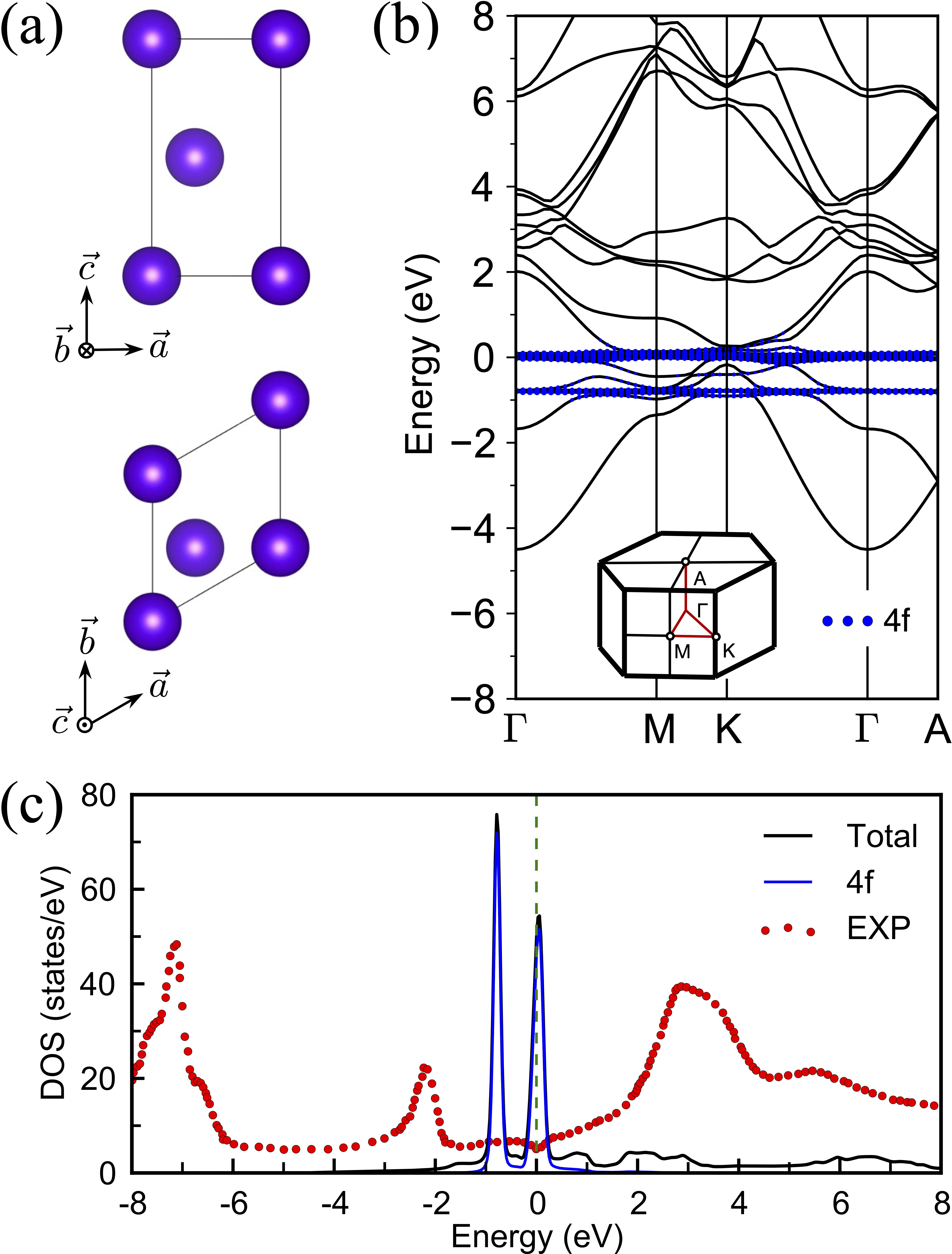}
\caption{\label{Figure 1}(a) Crystal structure of Tb in the hcp phase. (b) DFT band structures for Tb at ambient pressure.  The inset shows the first Brillouin zone and the high symmetry points. (c) DFT electron DOS for Tb at ambient pressure. The localized $4f$ bands are shown in blue, and the experimental spectra~\cite{Lang81JPFMP} are shown in red. Zero energy is set at the Fermi level.}
\end{figure}

{\it Results and Discussion --}
We first discuss the electronic structures in standard DFT calculations. At ambient pressure, Tb assumes a hexagonal close-packed (hcp) structure [Fig. 1(a)] with lattice parameters $a = 3.60$ {\AA} and $c = 5.72$ {\AA}~\cite{Kozlenko21PRM}. The DFT results in Fig. 1(b) show metallic band structures of Tb in the hcp phase, with the valence bands near the Fermi level predominantly contributed by $4f$ electrons. These $4f$ bands are nearly flat, lying in a narrow energy range of $\sim 0.2$ eV, suggesting strongly localized $4f$ orbitals as in an isolated Tb atom. Tb has valence electrons from open-shell $4f$, $5d$, and $6s$ orbitals. The $5d$ and $6s$ bands are more dispersive, lying in a much larger energy range over 2 eV. The orbitally resolved band structures on a zoomed-in energy scale are detailed in Fig. S1 of the Supplemental Material~\cite{SM}.

\begin{figure}
\includegraphics[width=0.9\columnwidth]{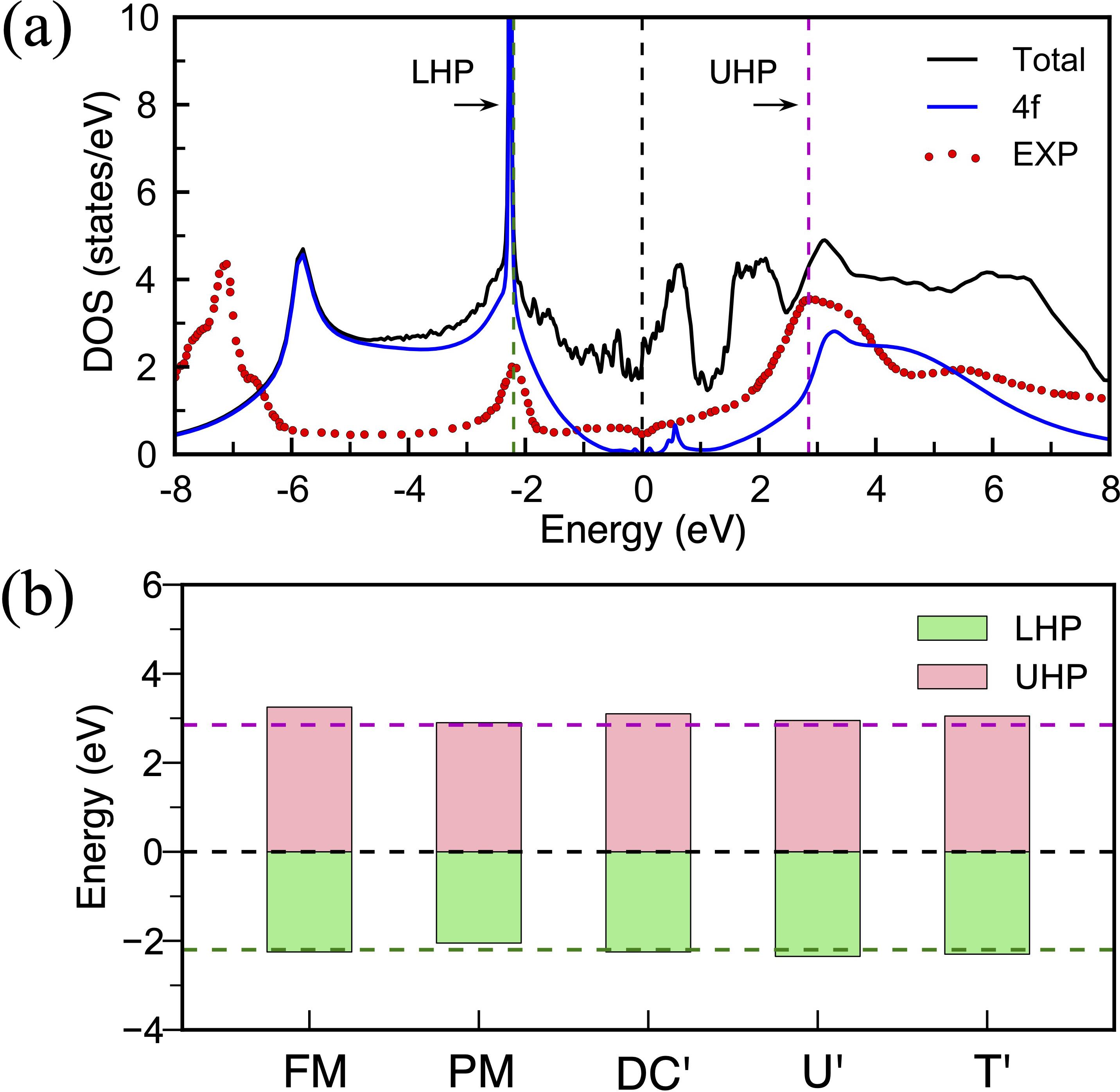} 
\caption{\label{Figure 2}(a) DFT+DMFT electron DOS for Tb at ambient pressure. The experimental spectra~\cite{Lang81JPFMP} are shown in red. The energies of the lower Hubbard band peak (LHP) and upper Hubbard band peak (UHP) are indicated by the green and pink dashed lines, respectively. Zero energy is set at the Fermi level. (b) The LHP and UHP energy locations in DFT+DMFT calculations for different computation schemes and parameters: Ferromagnetic (FM), paramagnetic (PM), a second double-counting scheme (DC$'$) \cite{Haule15PRL}, a lower Hubbard $U$ parameter ($U'$), and a lower temperature parameter ($T'$) (see further details in the Supplemental Material~\cite{SM}).}
\end{figure}

Figure 1(c) shows the DFT electron density of states (DOS) and the corresponding experimental spectra~\cite{Lang81JPFMP}. In the DFT calculations, two sharp peaks with $4f$ orbital character exist near the Fermi level, centered around $-0.8$ eV and 0.05 eV, respectively. These two peaks are split by the strong spin-orbit coupling (SOC) in this system~\cite{Lebegue06JPCM}. On the other hand, the experimental spectra exhibit three main peaks far from the Fermi level, with energies around $-7.1$ eV, $-2.2$ eV, and 2.85 eV, respectively (see also Fig. S2 in the Supplemental Material~\cite{SM}). The strong theory-experiment discrepancy necessitates a beyond-standard DFT description.

\begin{figure*}
\includegraphics[width=2\columnwidth]{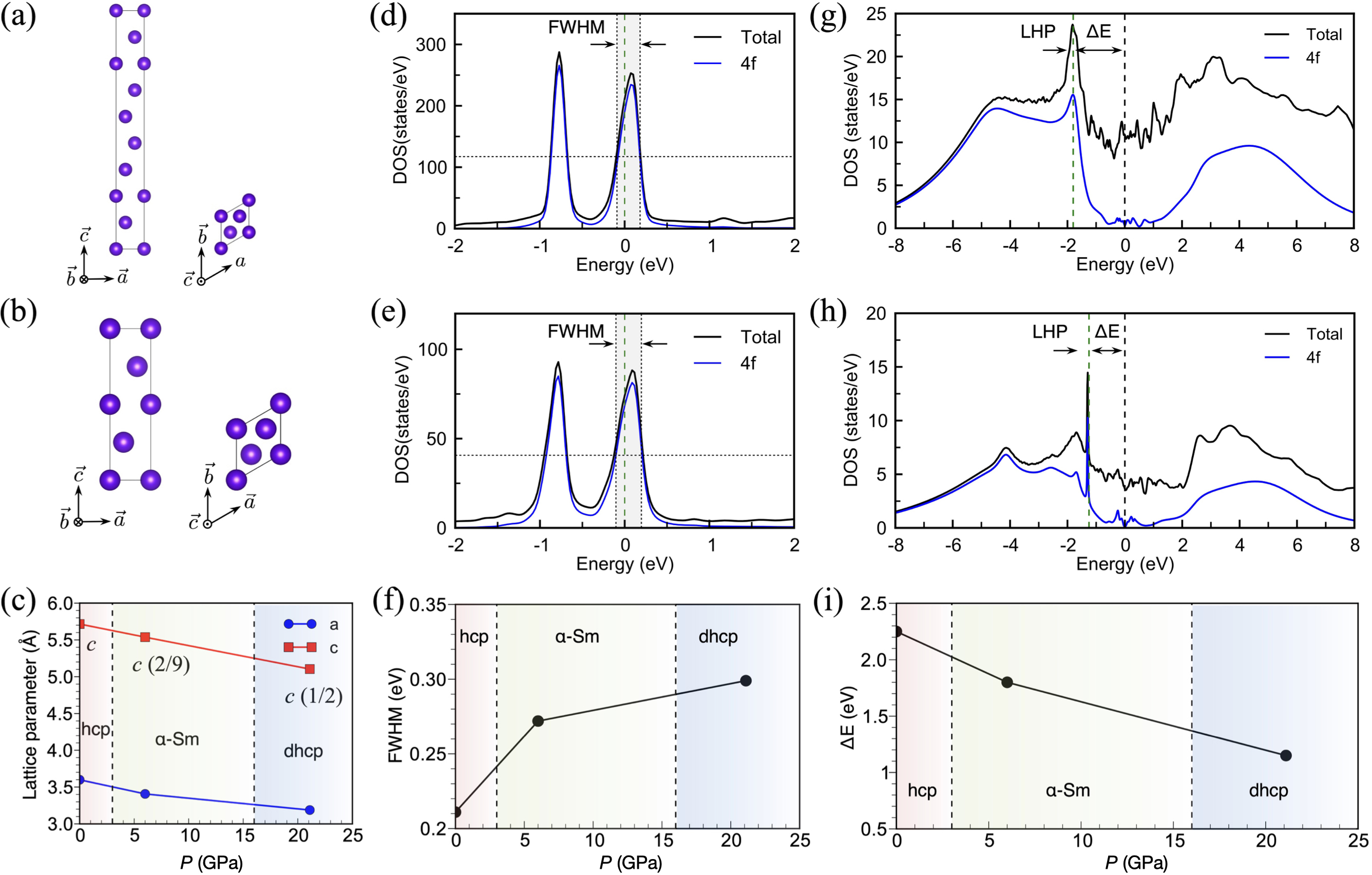}
\caption{\label{Figure 3}(a)-(b) Crystal structures of Tb in the $\alpha$-Sm and dhcp phases, respectively. (c) Experimental lattice parameters of Tb as a function of pressure. (d)-(e) DFT electron DOS for Tb in the $\alpha$-Sm and dhcp phases, respectively. (f) Full width at half maximum (FWHM) of the DFT bands [gray areas in (a)-(b)] as a function of pressure. (g)-(h) DFT+DMFT electron DOS for Tb in the $\alpha$-Sm and dhcp phases, respectively. (i) The LHP locations in DFT+DMFT calculations as a function of pressure. The Hubbard gap is effectively reduced by a pressure-enhanced charge hybridization. Zero energy is set at the Fermi level.}
\end{figure*}

We now turn our focus to the DFT+DMFT results. As shown in Fig. 2(a), the $4f$ states form a three-peak structure in the DFT+DMFT calculations, which agrees quantitatively well with the experimental spectra. In particular, the two peaks located around $-2$ eV and 3 eV are reminiscent of the lower Hubbard band peak (LHP) and upper Hubbard band peak (UHP) in a Hubbard model, respectively. The formation of the LHP and UHP manifests strong correlation effects due to $4f$-orbital on-site Coulomb interactions. The theory-experiment agreement in the electron DOS represents the most significant result of our work, highlighting the importance of a fully {\it ab initio} treatment of correlation effects. On the other hand, the third peak around $-7$ eV, along with other features at deeper binding energies, is likely caused by atomic multiplet effects~\cite{Lang81JPFMP} and is not fully resolved in the current DFT+DMFT calculations. These states at deeper binding energies are not expected to play any role in the transport or magnetic properties of the system.

The theoretical spectra in Fig. 2(a) are obtained using a Hubbard interaction $U = 6$ eV and a Hund's coupling $J = 0.7$ eV on the $4f$ orbitals of Tb in the hcp phase. The (screened) $U$ value is compatible with the first-principles value from linear response calculations~\cite{Logan24JPCM}. Unlike the Hubbard density-density interaction, it is more difficult to screen the Hund's exchange originating from higher-order multipole expansions of the atomic wavefunctions. Therefore, the $J$ value is only slightly reduced from the expected Hund's coupling in the atomic limit (typically $\sim 1.0$ eV)~\cite{van1988electron}. Below, we systematically examine the DFT+DMFT results and show that the LHP and UHP spectra are robust against minor variations in the calculation schemes and interaction parameters.

Figure 2(b) displays the energy locations of the LHP and UHP, respectively indicated by the horizontal dashed green and red lines. The exact input parameters and the resulting electron DOS in the DFT+DMFT calculations for Fig. 2(b) are given in Fig. S3 and the Supplemental Material~\cite{SM}. The energy separation between the LHP and UHP (i.e. the Hubbard gap) is roughly 5.0 eV. As seen in Fig. 2(b), the effective Hubbard gap is slightly enlarged in the ferromagnetic (FM) calculation compared to the paramagnetic (PM) case. In addition to magnetism, we have tested different double-counting schemes (which remove the interaction effects already included in the DFT exchange functional) and interaction parameters. As also shown in Fig. 2(b), the Hubbard gap remains essentially unchanged when we use a second double-counting scheme (DC$'$) or a slightly reduced Hubbard $U$ value. Finally, when we slightly reduced the temperature ($T'$), the Hubbard gap also basically remains the same. Overall, the variations in the energies of the LHP and UHP are within 0.15 eV and 0.40 eV from the experimental spectra, respectively. These results demonstrate the strength and stability of our fully charge self-consistent DFT+DMFT calculations in describing the correlated electronic structures of Tb. 

After achieving quantitative agreement in the hcp phase, we next apply the same methodology to study the higher-pressure $\alpha$-Sm and dhcp phases [see Figs. 3(a) and 3(b)]. Specifically, elemental Tb undergoes a structural transition from hcp to $\alpha$-Sm at $\sim 2.5$–3.5 GPa, and to the dhcp phase at $\sim 16$ GPa. Additional structural transitions can occur at even higher pressures~\cite{Matthew23JMMM, Kozlenko21PRM} but are not considered here. The hcp $\rightarrow$ $\alpha$-Sm $\rightarrow$ dhcp transitions exhibit a progressive change in the local Tb crystal environment, accompanied by moderate lattice parameter contractions [see Fig. 3(c)]. The DFT electron DOS for the $\alpha$-Sm and dhcp phases [respectively shown in Figs. 3(d) and 3(e)] display two clear peaks of $4f$ states split by strong SOC, which is similar to the hcp case discussed in Fig. 1(c). The full width at half maximum (FWHM) of the DFT band near the Fermi level is shown in Fig. 3(f). The peak width gradually increases from 0.21 eV in the hcp phase to 0.27 eV in the $\alpha$-Sm phase, and eventually to 0.30 eV in the dhcp phase. This effective band broadening indicates a pressure-enhanced electron kinetic energy or charge delocalization effect.

The DFT+DMFT results for the $\alpha$-Sm and dhcp phases are shown in Figs. 3(g) and 3(h), respectively. The 4$f$ states in the electron DOS move away from the Fermi level and form a three-peak structure consisting of the LHP and UHP, as well as a third peak at deeper binding energies associated with multiplet structures. These results are qualitatively similar to the hcp case discussed in Fig. 2(a), indicating that the electronic structures of all Tb's high-pressure phases remain strongly correlated and go beyond standard DFT descriptions. Interestingly, the energy separation between the LHP and the Fermi level [denoted by $\Delta E$ in Fig. 3(i)] monotonically decreases under pressure, changing from 2.25 eV in the hcp phase to 1.80 eV in the $\alpha$-Sm phase, and eventually to 1.15 eV in the dhcp phase. Therefore, the Coulomb interactions are effectively weakened by enhanced charge screening. This is consistent with a previous work on determining the effective Hubbard $U$ of Tb by first-principles linear-response  calculations~\cite{Logan24JPCM}. The result demonstrates the tunability of correlation effect by high pressure, which alters the competition between kinetic energy and Coulomb interaction. The evolution of the Hubbard gap is further discussed in Figs. S4 and S5 in the Supplemental Material~\cite{SM}.
The resulting DFT+DMFT electron DOS serve as predictions for benchmarking future spectroscopic experiments. 

\begin{figure}
\includegraphics[width=0.85\columnwidth]{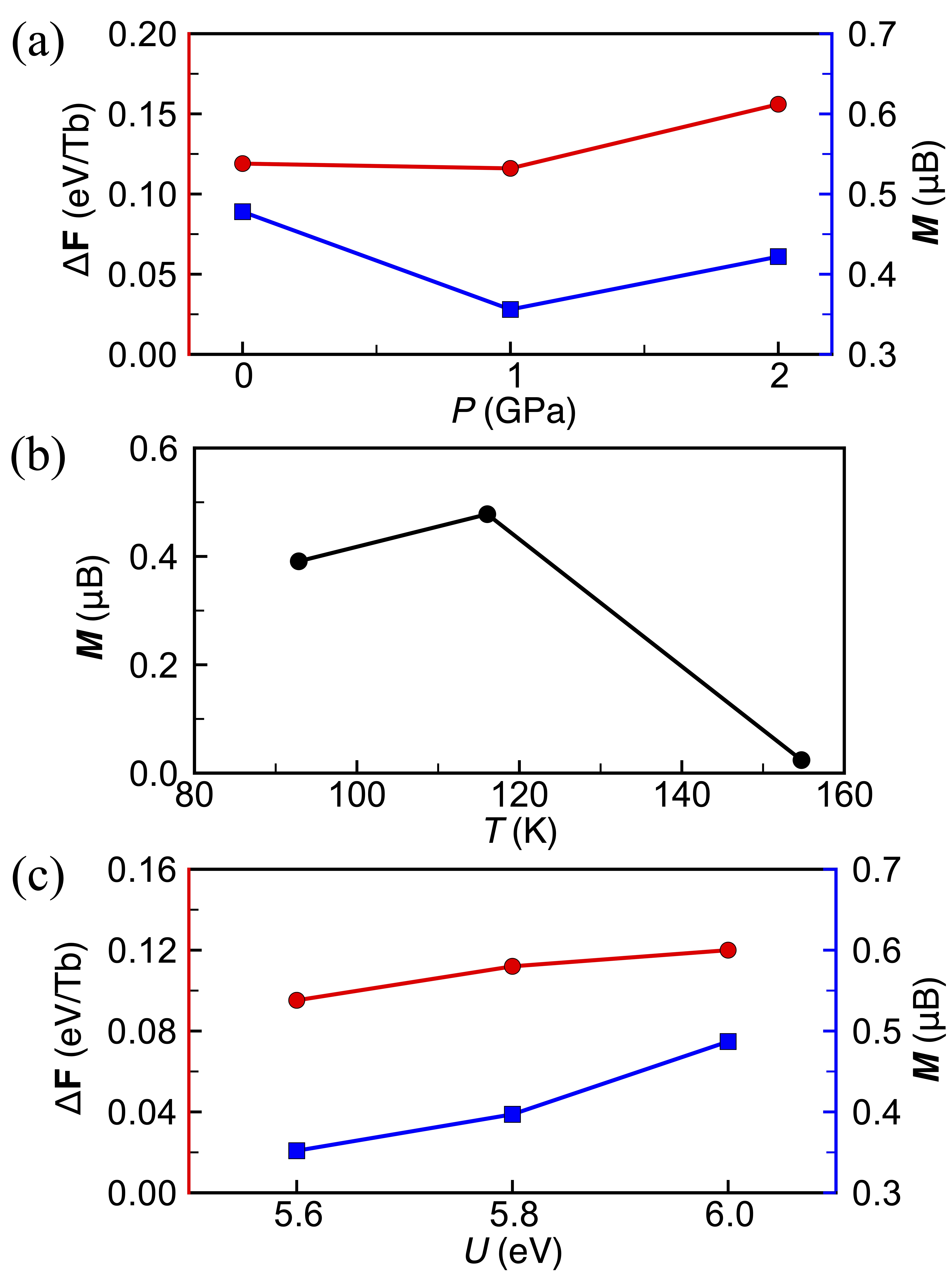}
\caption{\label{Figure 4}DFT+DMFT calculations of the relative free energy ($\Delta F \equiv F_{PM}-F_{FM}$, red curves) and the FM magnetic moment ($M$, blue curves) of Tb as a function of (a) pressure ($P$), (b) temperature ($T$), and (c) on-site Hubbard parameter $U$.}
\end{figure}

%Aside from the modulation of strong correlations by high pressure, the complex magnetisms reported in Tb depending on pressure and temperature parameters has given rise to another research topic for both experimental and theoretical interests \cite{Matthew23JMMM}. In particular, at ambient pressure, elemental Tb undergoes a series of magnetic transitions from paramagentism (PM) to helical antiferromagnetic ordering and then to ferromagnetism (FM) as temperature goes down, where the helical antiferromagnetic state only exists in a narrow temperature window from 230 to 220 K \cite{Koehler63JAP, Koehler65JAP, Dietrich67PR}. Furthermore, as pressure goes up, the ferromagnetic ground state at low temperatures may transform to several candidates of interlayer antiferromagnetic orderings \cite{Kozlenko21PRM, Matthew23JMMM}. To explore such as a complex magnetic phase diagram via first-principles calculations should require two crucial conditions. First, the strongly correlated nature of the valence electronic states of Tb has to be well described by the calculation methods at a quantitative level, as shown in our DFT+DMFT calculations on the hcp phase of Tb. Second, the energetic differentiation of the calculation methods should be accurate enough to capture the delicate differences between complex magnetic orderings.

Finally, we discuss the use of DFT+DMFT to study Tb's magnetic properties at low pressures and temperatures, as well as their dependence on the calculation parameters of pressure ($P$), temperature ($T$), and Hubbard interaction ($U$). As shown in Fig. 4(a), the relative free energy between the PM and FM states ($\Delta F = F_{PM} - F_{FM}$) remains positive between 0 and 2 GPa, indicating that the hcp phase favors long-range FM order at low temperature, which agrees with experiments~\cite{Koehler63JAP, Koehler65JAP, Dietrich67PR, Kozlenko21PRM, Matthew23JMMM}. On the other hand, when the simulation temperature $T$ increases from 92.8 K to 154.7 K, the magnetic moment $M$ (defined in the Supplemental Material~\cite{SM}) or the FM order parameter is suppressed to essentially zero [see Fig. 4(b)], indicating a temperature-induced FM to PM transition, again agreeing with experiments. We note that the experimental transition temperature to the FM phase is at 221 K. The theory-experiment temperature discrepancy corresponds to a small energy scale of 0.006 eV, which may be attributed to uncertainties in the interaction parameters and/or crystal structures, or small errors introduced in the impurity solvers.
Finally, when the Hubbard $U$ increases from 5.6 eV to 6.0 eV, both $\Delta F$ and $M$ increase accordingly [see Fig. 4(c)], indicating an enhanced tendency towards FM order due to Coulomb interactions~\cite{Logan24JPCM}. These results demonstrate that, in addition to describing the electronic structures, DFT+DMFT also captures the magnetic behavior of Tb reasonably well at low pressures and temperatures.

{\it Conclusion --}
We have performed the first fully charge self-consistent DFT+DMFT calculations to study the electronic structures and magnetism of Tb under pressure. We have obtained quantitative theory-experiment agreement in the electron density of states (DOS), demonstrating the capabilities of the employed method in simulating correlated electrons from first principles. Our results can be used to benchmark different numerical techniques and serve as predictions for future spectroscopic measurements on the high-pressure phases of Tb. In principle, the methodology of this work can be broadly employed to explore the correlated electronic structures of other heavy lanthanide and rare-earth materials under pressure. The DFT+DMFT results also correctly described the tendency toward ferromagnetism in the low-pressure hcp phase of Tb, as well as the suppression of FM order with increasing temperature. Meanwhile, AFM orders in the hcp or the higher-pressure $\alpha$-Sm and dhcp phases remain difficult to simulate directly in the current DFT+DMFT implementation. These simulations potentially require DFT supercells and a cluster extension of DMFT beyond a single quantum impurity. These challenging calculations would be critical in future studies to provide further insights into the complex magnetic phase diagrams of heavy lanthanides, where the underlying spin configurations are not fully understood at high pressures.

\begin{acknowledgments}
{\it Acknowledgments --}
This work is supported by the U.S. Department of Energy (DOE) Basic Energy Sciences Program under Award No. DE-SC0023268. 
This research used resources of the National Energy Research
Scientific Computing Center, a U.S. DOE Office of Science User Facility
supported under Contract No. DE-AC02-05CH11231 using NERSC Award BES-ERCAP0033090.
\end{acknowledgments}

%\bibliographystyle{apsrev4-2}
\bibliography{2024_Tb}


%\clearpage

\appendix{

\renewcommand{\thefigure}{A\arabic{figure}}
\setcounter{figure}{0}

\section{APPENDIX}

\section{I. Fully Charge Self-Consistent DFT+DMFT Calculation}

Figure A1 shows a schematic workflow chart of a fully charge self-consistent DFT+DMFT calculation, which consists of two major self-consistent iteration cycles at both the DMFT level~\cite{Kotliar96RMP, Lichtenstein01PRL, Lechermann06PRB, Kotliar06RMP, Amadon08PRB} and the DFT+DMFT charge level. The entire calculation process can be illustrated in three major steps:

\begin{enumerate}[label=\arabic*.]
\item 
The electronic structure of the material system under study is first obtained via standard DFT calculations, from which the tight-binding parameters of the lattice Hamiltonian can be derived by projecting the band structures to local orbitals. 
    
\item 
Within the DMFT cycle, an interacting lattice Hamiltonian is mapped onto a quantum impurity model, where an impurity site hybridizes with a bath environment. This embedding procedure naturally leads to a frequency-dependent self-energy for the impurity, which can be solved exactly by quantum Monte Carlo. A key quantity for achieving self-consistency in DMFT is the hybridization function ($\Delta$), which describes the correlation between the impurity site and the electron bath. This function mimics a dynamical mean-field, parameterized by the hopping terms between the impurity and the bath, as well as the site energies of the bath environment. Finally, the impurity self-energy is approximated as the self-energy of the original lattice problem. The process repeats until the self-energy (or Green's function) converges.
	
\item 
The charge density ($\rho$) of the correlated local orbitals from DMFT is fed back into DFT electronic structure calculations, which provide a new tight-binding Hamiltonian for the next DMFT cycle. The entire process then repeats until the charge density from both DFT and DMFT converges, leading to fully charge self-consistent DFT+DMFT {\it ab initio} simulations.

\end{enumerate}

\begin{figure}
\includegraphics[width=0.85\columnwidth]{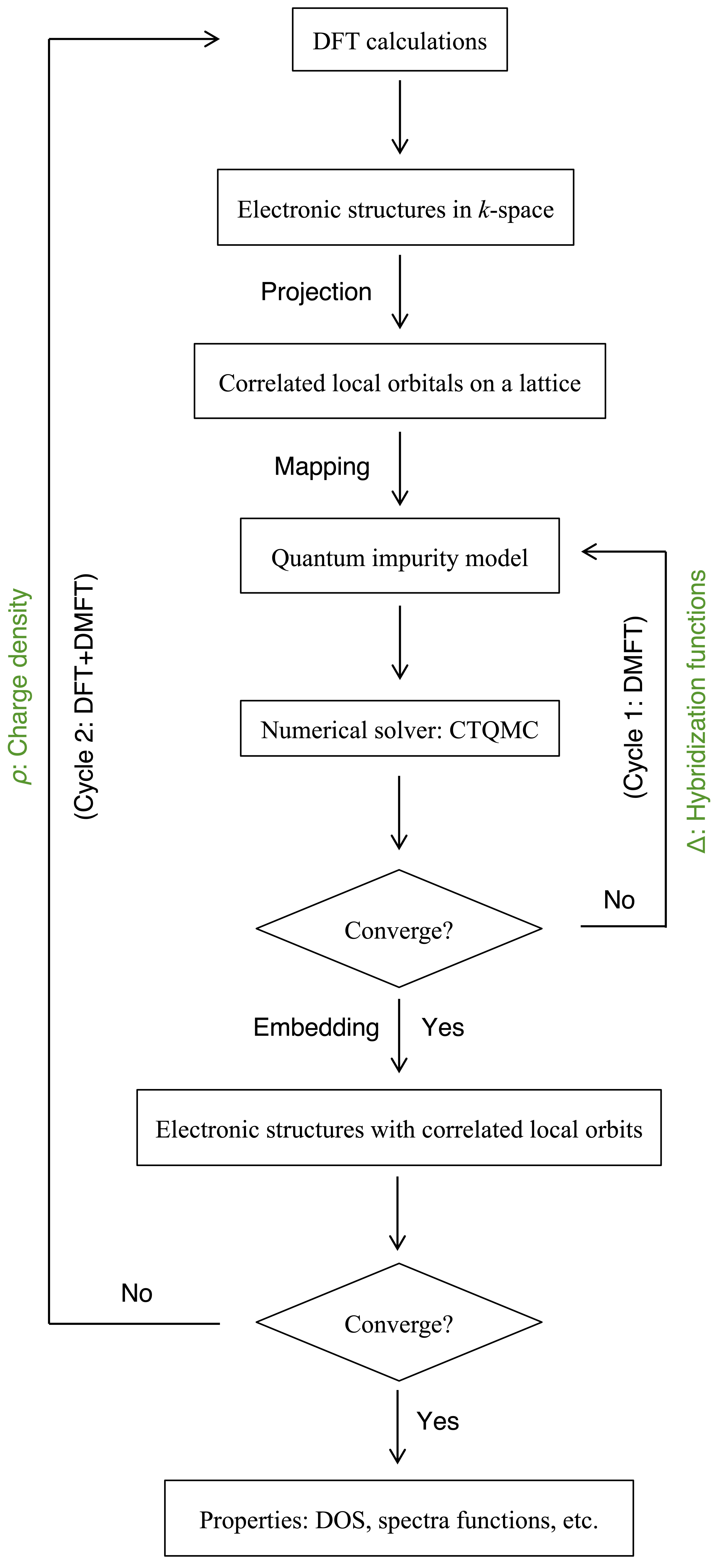}
\caption{\label{Figure A1} Schematic workflow chart of a fully charge self-consistent DFT+DMFT calculation.}
\end{figure}

\begin{figure*}
\includegraphics[width=1.95\columnwidth]{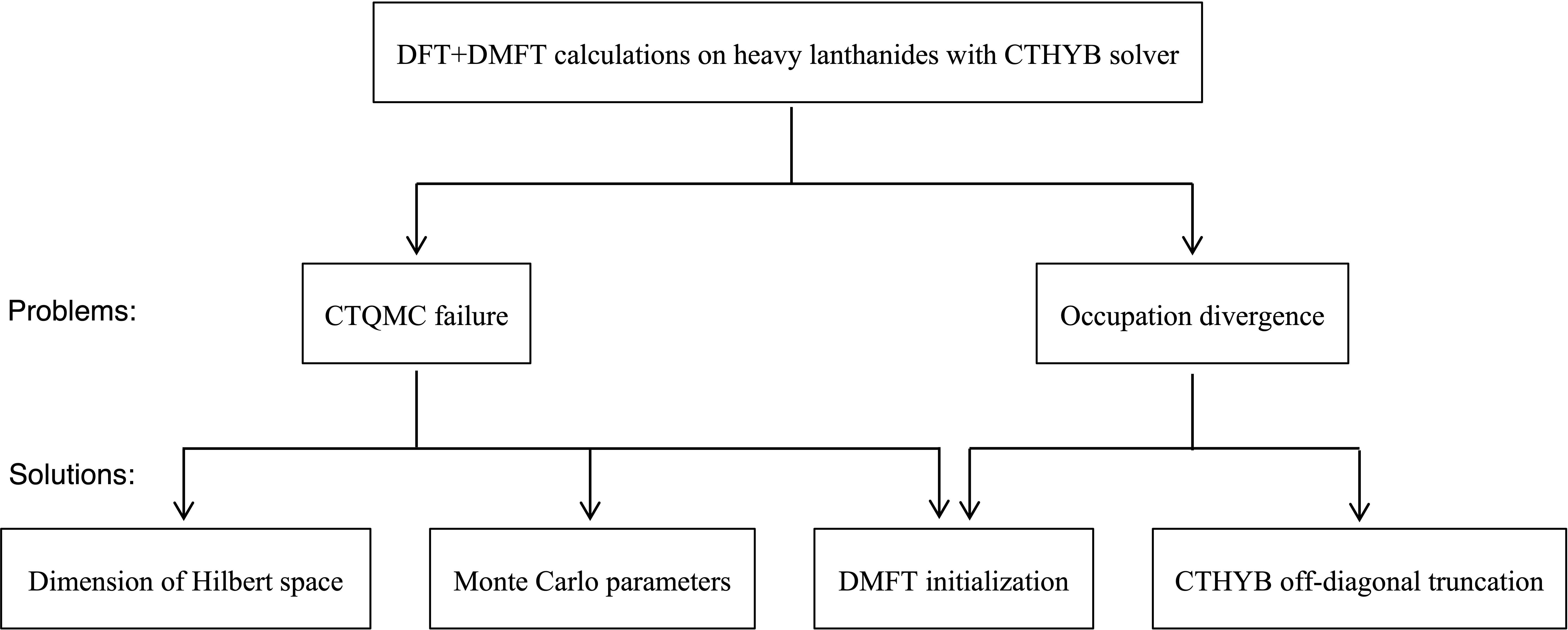}
\caption{\label{Figure A2} Schematic chart of the problems and solutions for achieving stable and convergent DFT+DMFT calculations on heavy lanthanides.}
\end{figure*}

\section{II. Problems and Solutions for Convergent DFT+DMFT Calculations in Heavy Lanthanides}

When performing DFT+DMFT calculations on Tb (and other heavy lanthanides), we found that it can be challenging to reach stable and convergent solutions using the EDMFTF software~\cite{Haule10PRB} with the CTQMC (continuous-time quantum Monte Carlo) solver employing CTHYB (hybridization expansion)~\cite{Werner06PRL, Haule07PRB, Gull11RMP}. Below, we divide the problems we encountered into two categories, describe the possible reasons for their occurrence, and provide solutions based on systematic and comprehensive tests of EDMFTF. The problems and solutions are also summarized in the schematic chart in Fig. A2.

The first problem is related to the CTQMC solver:

\begin{enumerate}[label=\arabic*.]
\item
One issue in simulating Tb and other heavy lanthanides is that the Hilbert space dimension for the $4f$ orbitals is much larger (compared to light lanthanides or $d$-electrons in transition metals). For example, in cerium (Ce), the atomic configuration space typically includes $f^0$, $f^1$, $f^2$, and $f^3$, leading to a maximal atomic matrix size of 41 (as used in the EDMFTF software). In the case of Tb, however, the atomic configurations of $f^7$, $f^8$, $f^9$, and $f^{10}$ are required, resulting in a maximal atomic matrix size of 327 (as used in the EDMFTF software). Therefore, a sufficiently large dimension for the atomic configuration space must be carefully specified in the software.

\item
The stability of DMFT calculations is sensitive to the input parameters for the Monte Carlo simulations. To address this issue in EDMFTF, we adjust the parameters of the Monte Carlo steps in the range of $10^7$ to $10^8$ (with tsample = 100) and the inverse temperature ($\beta$) in the range of 50 to 100 (or temperature $T$ between 116.1 K to 232.1 K) to achieve stable calculations.

\item 
The stability is also sensitive to the initialization of DMFT conditions. Important parameters to set include the nominal valence occupancy ($n_{f0}$), the double-counting scheme (DC), the projection energy window, and the initial magnetic state. To achieve stable DFT+DMFT results for Tb within the EDMFTF software, we typically use a nominal valence occupancy of 9, test different double-counting schemes (``nominal" or ``exact"), choose a projection window that is as narrow as possible but still encloses all $4f$ states, and test both ferromagnetic and paramagnetic configurations.

\end{enumerate}

The second problem is the divergence between the lattice occupation ($n_{\rm{lat}}$) and the impurity occupation ($n_{\rm{imp}}$) of the correlated atom, which is an important criterion for achieving convergence in DMFT calculations:

\begin{enumerate}[label=\arabic*.]
\item 
A ``bad" initialization of DMFT conditions can lead to stable but non-convergent calculation results, namely, a large discrepancy between $n_{\rm{lat}}$ and $n_{\rm{imp}}$. In this case, both the Green's functions and self-energies may become unphysical and should be monitored to prevent further issues.

\item
A truncation scheme for the off-diagonal terms in the hybridization functions is adopted in the quantum impurity solver. This is a conventional treatment for improving the sign problem when using CTHYB solvers, which, however, may lead to a small discrepancy between $n_{\rm{lat}}$ and $n_{\rm{imp}}$ (see, for example, Fig. S6 of the Supplemental Material~\cite{SM}). Adjustments to the local orbital basis may reduce this small discrepancy. However, the resolution of this issue is beyond the scope of the present work.
\end{enumerate}

After systematic and comprehensive tests, our DFT+DMFT calculations on Tb have reached stable convergence in the free energies, Green functions, and self-energies. These results are displayed in Fig. S6 of the Supplemental Material~\cite{SM}.	
}

\end{document}

%\preprint{}

\title{Supplemental Material\\
Terbium under High Pressure: First-Principles Dynamical Mean-Field Theory Study}


\author{Wenjun Ding}
\email{wding@uab.edu}
\affiliation{Department of Physics, University of Alabama at Birmingham, Birmingham, Alabama 35294, USA}
\author{Yogesh K. Vohra}
\affiliation{Department of Physics, University of Alabama at Birmingham, Birmingham, Alabama 35294, USA}
\author{Cheng-Chien Chen}
\email{chencc@uab.edu}
\affiliation{Department of Physics, University of Alabama at Birmingham, Birmingham, Alabama 35294, USA}

\date{\today}

%\keywords{density functional theory, dynamical mean-field theory, terbium, rare-earth metal, high pressure}

\maketitle

\noindent \textbf{1. Calculation methods and details}

The fully charge self-consistent density functional theory (DFT) plus dynamical mean-field theory~\cite{Kotliar96RMP, Lichtenstein01PRL, Lechermann06PRB, Kotliar06RMP, Amadon08PRB} (DFT+DMFT) calculations are performed within the EDMFTF software~\cite{Haule10PRB} interfaced with the WIEN2k package~\cite{Blaha20JCP}. In the DFT part of WIEN2k, the valence electrons and core electrons are described using the all-electron full-potential linearized augmented-plane-wave (LAPW) method, and the exchange and correlation functionals are treated with the local density approximation (LDA). The plane-wave basis parameter (RKmax) in the WIEN2k code is set to 8.5, and a $\Gamma$-centered $k$-mesh is chosen to be 20 $\times$ 20 $\times$ 10 (5000 $k$ points), 14 $\times$ 14 $\times$ 14 (3000 $k$ points), and 26 $\times$ 26 $\times$ 7 (5000 $k$ points) for the hcp phase, the $\alpha$-Sm phase, and the dhcp phase of Tb, respectively. In the DMFT part, all 4$f$ orbitals of Tb are treated as the correlated orbitals, with the on-site Coulomb repulsion parameter (Hubbard $U$) chosen to be 6.0 eV and the exchange coupling parameter (Hund's $J$) chosen to be 0.7 eV, which are reasonable values for the lanthanides~\cite{Haule05PRL, Amadon14PRB, Locht16PRB}. The quantum impurity model is solved with the continuous time quantum Monte Carlo (CTQMC) method~\cite{Werner06PRL, Haule07PRB, Gull11RMP} within the ``nominal" double counting scheme. The maximum entropy method, as coded in the EDMFTF software, is used for performing analytic continuation of the electron self-energy to the real-frequency axis. The density of states are calculated with 10000 $\Gamma$-centered $k$ points for all phases of Tb. The inverse temperature parameter for the CTQMC calculations is chosen to be $\beta = 100$ ($T = 116.1$ K), $\beta = 75$ ($T = 154.7$ K), and $\beta = 50$ ($T = 232.1$ K) for the hcp phase, the $\alpha$-Sm phase, and the dhcp phase, respectively. The DMFT cycles are performed for 10 iterations or more, until convergences are reached for the free energy, Green's functions, and self-energies, as shown in Fig. S6. It is noted that the small discrepancy between the lattice occupation ($n_{\rm{lat}}$) and the impurity occupation ($n_{\rm{imp}}$) in Fig. S6 is mainly caused by the off-diagonal term truncation method used in the impurity solver. 
In the hcp phase of Tb, both paramagnetic and ferromagnetic states have been studied. For initializing ferromagnetism, a step-progressive energy difference of 0.2 eV has been added to the impurity levels of the $J = 7/2$ states in the initial self-energy. The $\alpha$-Sm and dhcp phases of Tb have been studied for paramagnetism only.
\\

\noindent \textbf{2. Parameters used in Fig. 2(b) of the main text}

In obtaining Fig. 2(b) of the main text, a systematic study has been performed on examining the quantitative agreements between our DFT+DMFT calculations and experiments under the variation of major calculation schemes and parameters. As described above, the standard calculation scheme and parameters for the hcp phase of Tb include ferromagnetism, ``nominal" double-counting scheme, on-site Hubbard $U$ parameter of 6.0 eV, and temperature of $T=116.1$ K. One of these calculation scheme and parameters has been changed for each test, including: (1) paramagnetism, (2) ``exact" double-counting scheme~\cite{Haule15PRL}, (3) a smaller Hubbard $U$ parameter of 5.8 eV, and (4) a lower temperature parameter of $T=92.8$ K. The results are displayed in Fig. S3.
\\

\noindent \textbf{3. Definition of the magnetic ordering parameter in Fig. 4 of the main text}

The magnetic ordering parameter ($M$) discussed in Fig. 4 of the main text is defined as:
\begin{align}
    M = \sum_{J = 5/2, \; 7/2}{ \Delta n_J^{spin} \sqrt{J(J+1)}}, 
\end{align}
\begin{align}
    \Delta n_J = n_{J, \uparrow} - n_{J, \downarrow}.
\end{align}
Here, $\Delta n_J$ is the difference of electron occupation between spin-up and spin-down states for total angular moment $J$ (= 5/2 or 7/2).
\\

\noindent \textbf{4. Supplemental figures}

Additional figures discussed in the main text and in this document are given below.

\renewcommand{\thefigure}{S\arabic{figure}}

\begin{figure}[h]
\includegraphics[width=0.85\columnwidth]{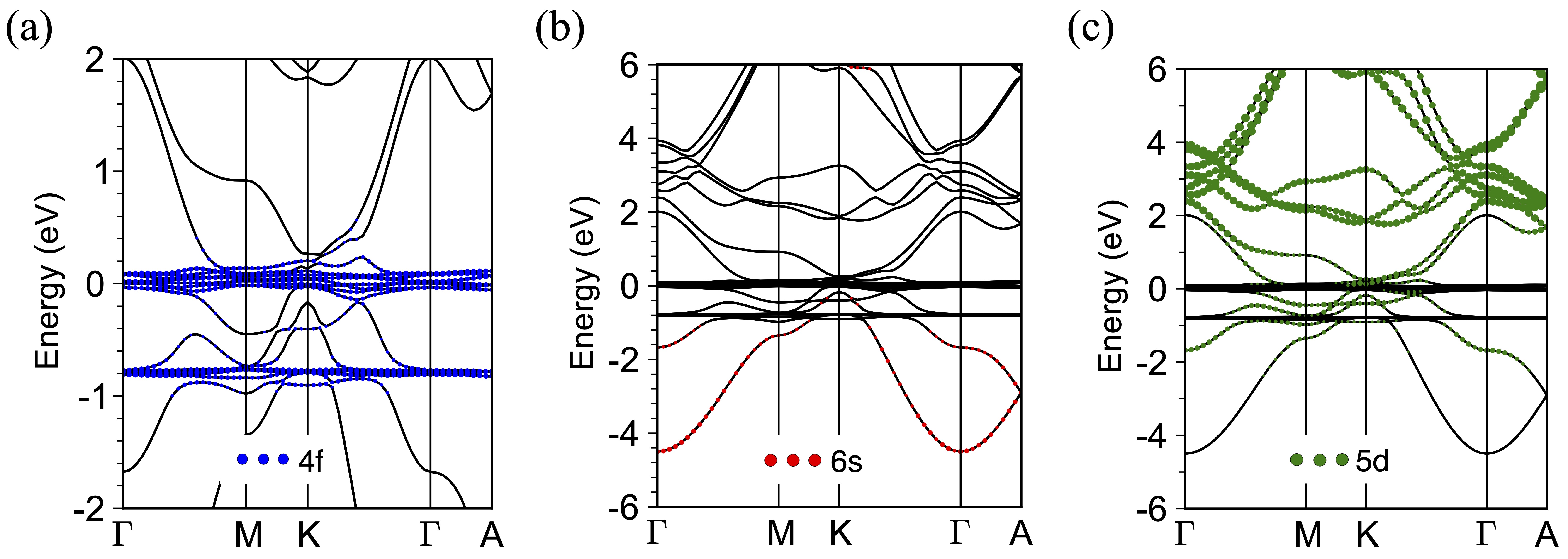}
\caption{\label{Figure S1} DFT band structures of Tb in the ambient-pressure hcp phase, with projections onto (a) $4f$ orbitals, (b) $6s$ orbitals, and (c) $5d$ orbitals, respectively. The projection intensities are represented by the circle sizes, colored in blue, red, and green, respectively. Zero energy is set at the Fermi level.}
\end{figure}

\begin{figure}
\includegraphics[width=0.85\columnwidth]{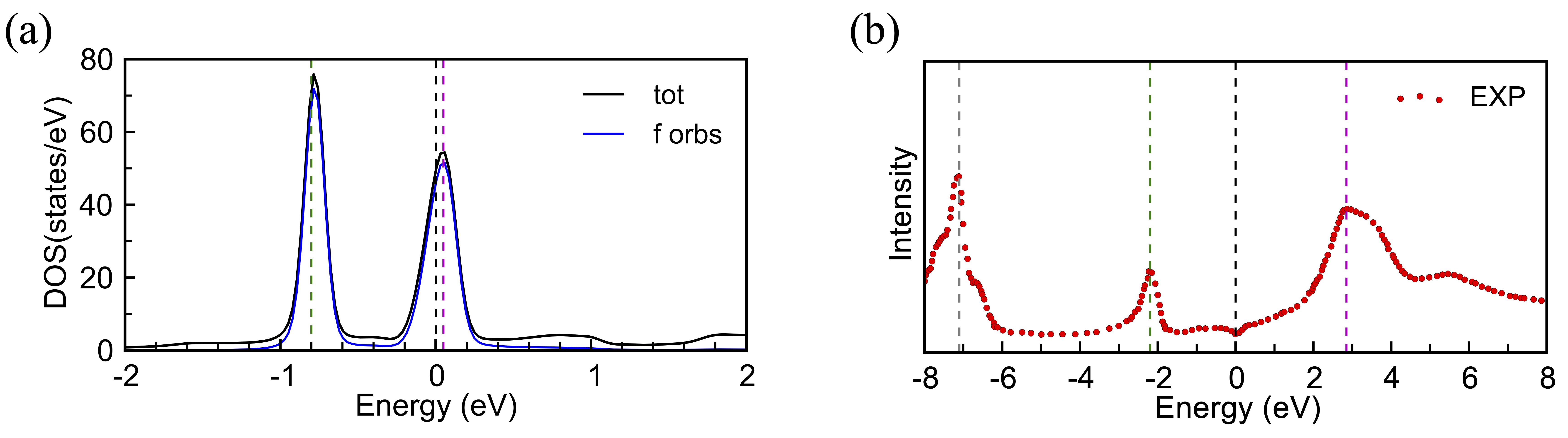}
	volume = {11},
\caption{\label{Figure S2} (a) DFT electron DOS of Tb in the ambient-pressure hcp phase. The total DOS and $4f$-orbital projected DOS are plotted in black and blue lines, respectively. The two peaks split by strong spin-orbit coupling (SOC) of $4f$ orbitals below and above the Fermi level are indicated by the green and pink lines, respectively. (b) Experimentally measured electron DOS of Tb in the ambient-pressure hcp phase~\cite{Lang81JPFMP}. The three major peaks contributed by $4f$ orbitals are labeled in gray, green, and pink lines, respectively. Zero energy is set at the Fermi level.}
\end{figure}

\begin{figure}
\includegraphics[width=0.85\columnwidth]{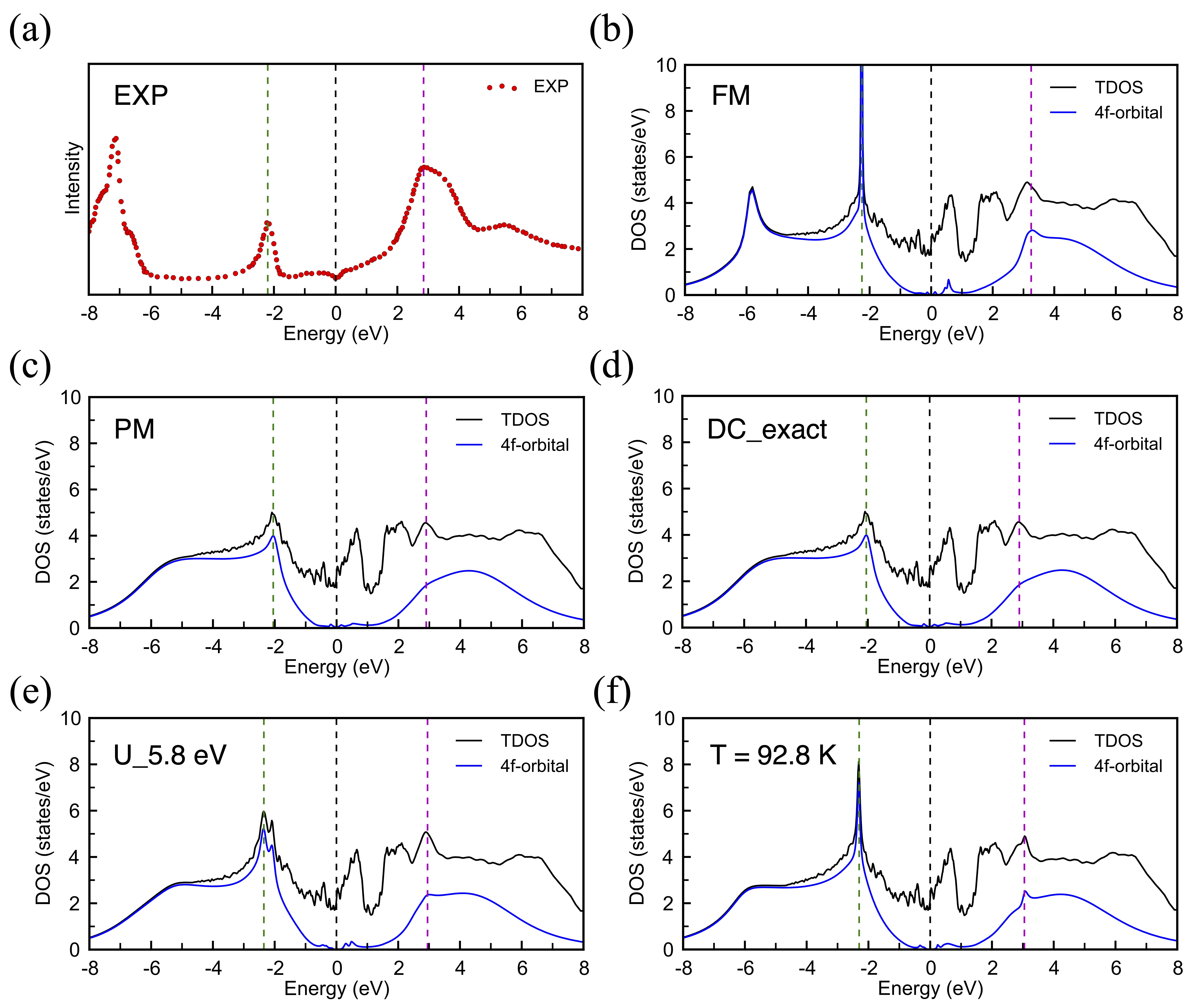}
\caption{\label{Figure S3} (a) Experimentally measured electron DOS of Tb in the ambient-pressure hcp phase~\cite{Lang81JPFMP}. (b) DFT+DMFT electron DOS within the standard setting of a ferromagnetic (FM) state, or under variations of the calculation schemes and parameters on (c) paramagnetic state (PM), (d) the ``exact" double-counting scheme (DC$_\textrm{exact}$), (e) a smaller Hubbard $U = 5.8$ eV, and (f) a lower temperature $T=92.8$ K. See more detailed discussions in Sec. 2 of this document. Here, the lower Hubbard band peak (LHP) and upper Hubbard band peak (UHP) are labelled in green and pink lines, respectively. Zero energy is set at the Fermi level.}
\end{figure}

\begin{figure}
\includegraphics[width=0.85\columnwidth]{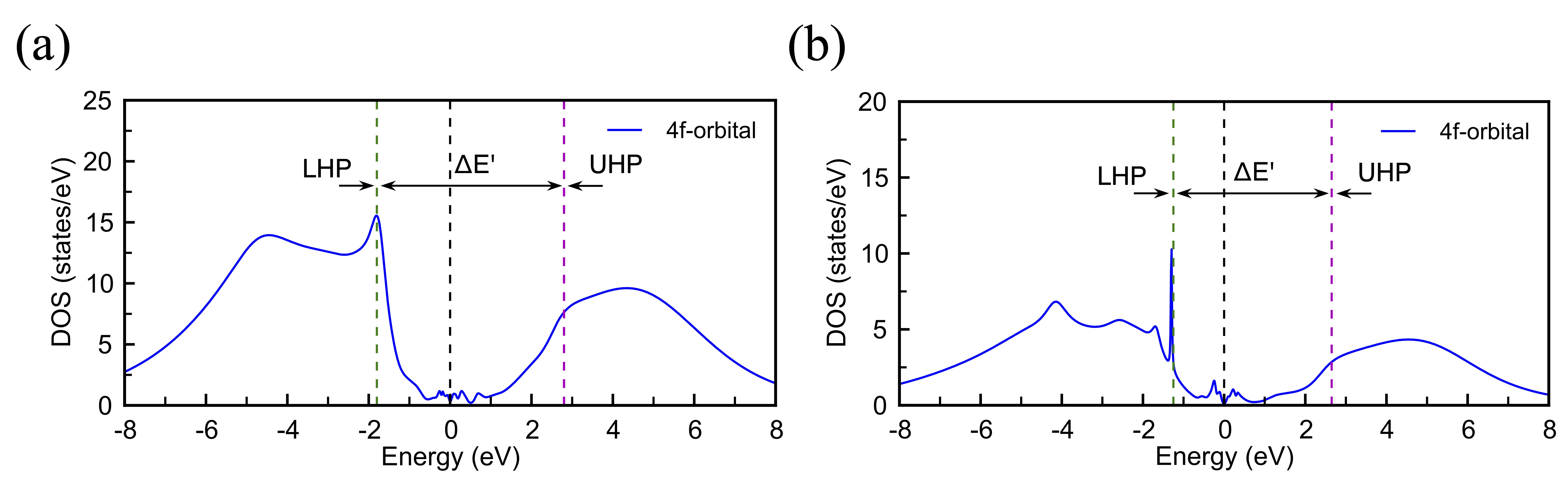}
\caption{\label{Figure S4} DFT+DMFT electron DOS for (a) the $\alpha$-Sm phase, and (b) the dhcp phase. The LHP and UHP are labelled in green and pink lines, respectively. The Hubbard gap or the energy difference ($\Delta \rm{E'}$) between the LHP and UHP is labelled with arrows. Zero energy is set at the Fermi level.}
\end{figure}

\begin{figure}
\includegraphics[width=0.85\columnwidth]{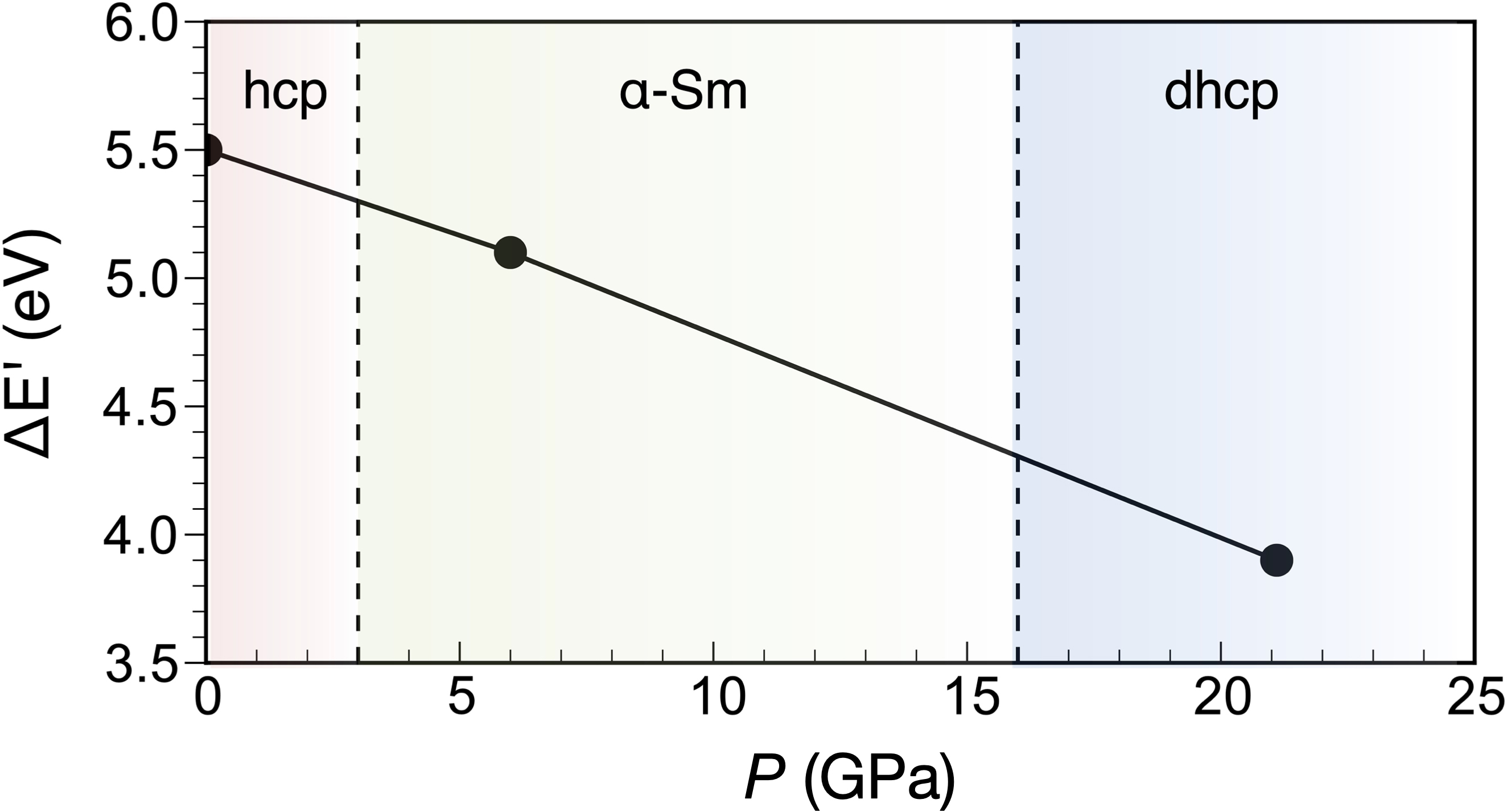}
\caption{\label{Figure S5} The Hubbard gap or the energy difference ($\Delta \rm{E'}$) between the LHP and UHP (defined in Fig. S4) as a function of pressure.}
\end{figure}

\begin{figure}[h]
\includegraphics[width=0.85\columnwidth]{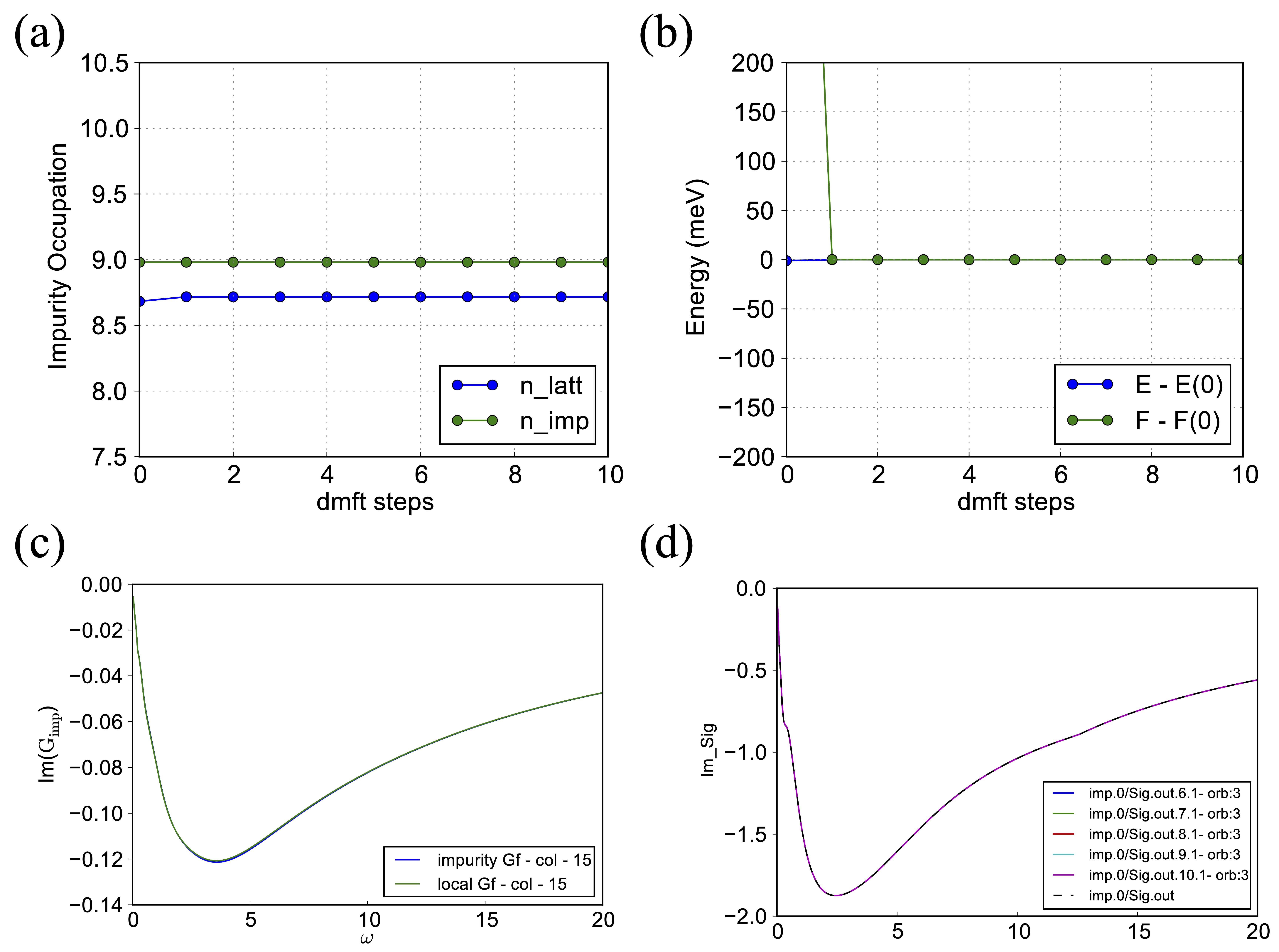}
\caption{\label{Figure S6} DFT+DMFT convergence tests on (a) the impurity- and lattice-site occupations, (b) the total and free energies, (c) the Green's functions, and (d) the self-energies. The calculations are performed using the ambient-pressure hcp structure of Tb.}
\end{figure}

\clearpage

\bibliographystyle{apsrev4-2}
\bibliography{2024_Tb_SM}